\documentclass[reprint,superscriptaddress]{revtex4-1}
\usepackage{amsmath}
\usepackage{graphicx}
\usepackage{bm}
\usepackage{natbib}
\def\apjl{Astrophys. J. Lett.}

\def\mnras{Mon.Not.Roy.As.Soc.}

\def\ltsima{$\; \buildrel < \over \sim \;$}
\def\simlt{\lower.5ex\hbox{\ltsima}}
\def\gtsima{$\; \buildrel > \over \sim \;$}
\def\simgt{\lower.5ex\hbox{\gtsima}}
\let\sec=\section

\def\[{\begin{equation}}
\def\]{\end{equation}}

\def\m@th{\mathsurround=0pt }
\def\eqalign#1{\null\,\vcenter{\openup1\jot \m@th
 \ialign{\strut\hfil$\displaystyle{##}$&$\displaystyle{{}##}$\hfil
 \crcr#1\crcr}}\,}

\begin{document}
\title{Clipping the Cosmos II: Cosmological information from non-linear scales}

\author{Fergus Simpson}
\email{frgs@roe.ac.uk}
\affiliation{SUPA, Institute for Astronomy, University of Edinburgh, Royal Observatory, Blackford Hill, Edinburgh EH9 3HJ, UK}

\author{Alan  F. Heavens}
\affiliation{Imperial Centre for Inference and Cosmology, Department of Physics, Imperial College,
Blackett Laboratory, Prince Consort Road, London SW7 2AZ, UK}

\author{Catherine Heymans}
\affiliation{SUPA, Institute for Astronomy, University of Edinburgh, Royal Observatory, Blackford Hill, Edinburgh EH9 3HJ, UK}

\newcommand{\ud}{\mathrm{d}}
\newcommand{\fpe}{f_\perp}
\newcommand{\fpa}{f_\parallel}
\newcommand{\om}{\Omega_m}
\newcommand{\dmax}{{\delta^{\rm{max}}}}
\newcommand{\kmax}{{k_{\rm{max}}}}
\newcommand{\rhomax}{{\rho_{\rm{max}}}}
\newcommand{\Veff}{{V_{\rm{eff}}}}
\newcommand{\eff}{{\rm{eff}}}
\newcommand{\lcdm}{$\Lambda$CDM }
\newcommand{\mpc}{ \, \rm{ Mpc}}
\newcommand{\hmpc}{ \, h \mathrm{ Mpc}^{-1}}
\newcommand{\hmpcv}{ \, h^3 \rm{ Mpc}^{-3}}
\newcommand{\hinvmpc}{ \, h^{-1} \rm{ Mpc}}
\newcommand{\tripleint}{\int \! \! \int \! \! \int}
\newcommand{\kfund}{k_f}
\newcommand{\fm}{f_m}
\newcommand{\dint}{\int  \!\!  \int}
\newcommand{\da}{\delta_a}
\newcommand{\db}{\delta_b}
\newcommand{\dc}{\delta_c}
\newcommand{\dca}{\delta_{ca}}
\newcommand{\dcb}{\delta_{cb}}
\newcommand{\dsi}{\delta_{si}}
\newcommand{\dclip}{\delta_{c0}}
\newcommand{\dThresh}{\delta_{0}}
\newcommand{\deltaX}{\delta^{(X)}}
\newcommand{\rh}{{\rho}(r)}

\date{\today}

\begin{abstract}
We present a method for suppressing contributions from higher-order terms in perturbation theory, greatly increasing the amount of information which may be extracted from the matter power spectrum. In an evolved cosmological density field the highest density regions are responsible for the bulk of the nonlinear power. By suitably down-weighting these problematic regions we find that the one- and two-loop terms are typically reduced in amplitude by $\sim 70\%$ and $\sim 95\%$ respectively, relative to the linear power spectrum. This greatly facilitates modelling the shape of the galaxy power spectrum, potentially increasing the number of useful Fourier modes by more than two orders of magnitude. We provide a demonstration of how this technique allows the galaxy bias and the amplitude of linear matter perturbations $\sigma_8$ to be determined from the power spectrum on conventionally nonlinear scales, $0.1<k<0.7 \hmpc$.
\end{abstract}

\maketitle

\section{Introduction}

The spatial distribution of galaxies in the Universe encodes a wealth of information such as the abundance of dark matter, the mass of the neutrino, the conditions of the early Universe, and the laws of gravity. However reliably decoding this pattern on all but the greatest scales has proved highly challenging due to the nonlinear physics associated with gravitational collapse. As a result, the vast majority of the available cosmological information from galaxy redshift surveys is discarded. Only the largest Fourier modes are included, typically $k \lesssim 0.1 \hmpc$, to ensure their behaviour can be well described by perturbation theory. Given that the total number of modes scales as $k_{\rm{max}}^3$, there is clearly great scope for improvement. 

Standard perturbation theory suffers from two significant limitations. In the nonlinear regime the magnitude of higher-order terms does not decay rapidly, meaning a large number of terms needs to be evaluated, and this becomes computationally prohibitive beyond second order. While efforts have been made to incorporate these higher-order terms \cite{RPTCrocceScoccimarro}, this does not circumvent the more fundamental limitation. Perturbation theory relies on the velocity field being curl-free and single valued, an assumption which breaks down for highly virialised structures. Instead of attempting to model the full nonlinear power spectrum, we can manipulate the density field prior to evaluating the power spectrum. This has the potential to ameliorate various issues associated with nonlinear structure formation. In an effort to recover linear clustering statistics on conventionally nonlinear scales, various local transformations of the density field have been proposed such as Gaussianisation \cite{1992MNRAS.254..315W}, the logarithmic field \cite{Neyrinck2009, 2011WangLog, 2012Carron}, and clipping \cite{SimpsonClip}. They all share a key characteristic which is the strong suppression of very high density regions. This penalises the most massive dark matter halos which dominate small scale clustering statistics, allowing the less evolved regions to contribute a greater proportion of the signal. 

Consider the vast ensemble of Hubble patches spanning the Universe, as suggested by the inflationary paradigm. By chance, a handful of these will consist of an extremely smooth distribution of matter, an acute case of cosmic variance. What are the observational consequences within these exceptional regions? Nonlinear structure formation is substantially weakened. Observers in such a patch (rare as they may be) would infer a false form of the primordial power spectrum. But they would also find the growth of cosmic structure considerably easier to model than typical observers at an equivalent epoch. This is simply because perturbation theory offers a much more faithful representation of clustering statistics when the perturbations are weaker.  However we need not restrict ourselves to thinking on such large scales. The vast majority of our own observable Universe lies far beyond the gravitational influence of a galaxy cluster. At present, around $99.9\%$ of the cosmic volume and approximately $90\%$ of galaxies lie further than $3 \mpc$ from any halo exceeding $10^{14}$ solar masses. Yet it is these most massive conglomerations which dominate clustering statistics on nonlinear scales, especially higher order moments of the density field. Looking in the gaps between these highly nonlinear lumps offers a glimpse into the linear regime, and therefore an easier path to model the distribution of cosmic matter. 

In this work we shall seek a prescription for preferentially selecting those moderately-dense regions whose statistical properties more closely match the predictions of cosmological perturbation theory. To achieve this, we shall focus specifically on the clipping method. One feature of clipping is that by volume the vast majority of the density field is left untouched, thereby offering beneficial noise properties. A further advantage, which we shall highlight in this work, is that the power spectrum associated with a clipped Gaussian random field is known exactly, and it is to a very good approximation directly proportional to the original power spectrum. This proves crucial in developing a theoretical model for the clipped power spectrum.  When clipping is applied, the amplitude of higher-order terms decay much more rapidly.  And it is the clipped regions, where the most massive halos lie, which most grossly violate the assumptions of perturbation theory.  Clipping may also prove advantageous in ameliorating the influence of baryons, which are notoriously difficult to model in cosmological simulations, since they are most disruptive in the highest-density environments. 

In \S  \ref{sec:pk_real} we review the clipping technique and summarise our model of the power spectrum for a clipped cosmological field.  We present the exact solution for the power spectrum of a clipped Gaussian field in  \S \ref{sec:spectrum}. More general hybrid fields are considered in \S \ref{sec:coeffs}, before considering cosmological fields in  \S \ref{sec:cosmo}. Our results are presented in  \S \ref{sec:results}. 

\section{Clipped Power Spectrum} \label{sec:pk_real}
When studying cosmological structure the field of interest is the fractional density perturbation, $\delta(x) \equiv \rho(x)/\bar{\rho} - 1$. This may be decomposed into a power series

\[ \label{eq:expansion}
\delta_m = \delta^{(1)} + \delta^{(2)} + \delta^{(3)} + \deltaX \, ,
\]

\noindent where $\delta^{(n)} = \mathcal{O}(\delta_G^n)$, and $\delta_G$ represents the primordial field which we shall assume to be a Gaussian Random Field (GRF). The residual term $\deltaX$ represents the difference between the true field and our approximation. This  incorporates a wide variety of sources, not only the omitted higher-order terms $\sum \delta^{(n)}$, but also the breakdown of perturbation theory, the influence of baryonic physics, and non-linear galaxy bias.  Each of these terms is most problematic when $\delta_m$ is large. In order to construct a continuous field $\delta(x)$ from a set of discrete objects, such as dark matter particles or galaxies, we are required to select a smoothing scale. Fluctuations on scales below this smoothing length will be neglected.  

\begin{figure}
\includegraphics[width=80mm]{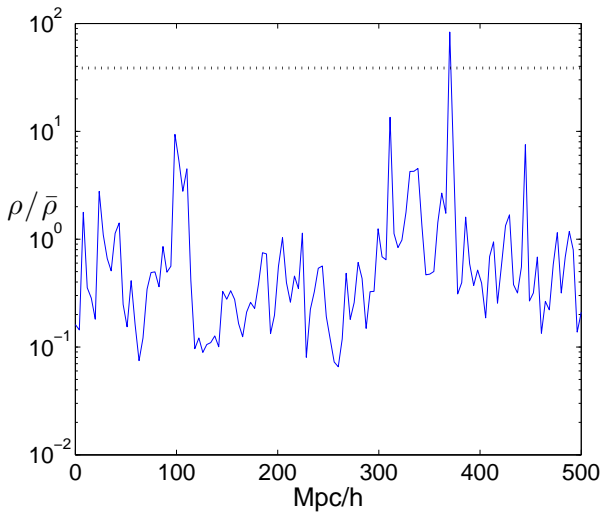}
\caption{An example of clipping applied to a one-dimensional skewer through a simulated cosmological density field at $z=0.$\label{fig:millClip}}
\end{figure}

In \citet{SimpsonClip} it was found that upon clipping, the theoretical relationship linking the power spectrum and bispectrum of dark matter and galaxies became valid across a much wider range of scales. Making use of this relationship out to $\kmax = 0.7 \hmpc$ enabled the galaxy bias to be recovered with high precision from a mock catalogue. However the functional form of the clipped power spectrum was at the time unknown. Establishing a suitable model for the clipped power spectrum therefore serves as the focus of this work.  Following \citet{SimpsonClip}, we generate the clipped field $\delta_c (x)$  by applying a threshold value $\delta_0$ such that

\[ \label{eq:clip}
\eqalign{
\delta_c (x)  &=  \delta_0 \qquad \, \, \, \,  (\delta_m (x) > \delta_0 )  \cr
\delta_c (x)  &=  \delta_m(x) \quad   (\delta_m (x) \leq  \delta_0 )  \cr
}
\]
which we may decompose into clipped component fields
\[ \label{eq:components}
\delta_c (x)  = \delta_c^{(1)} + \delta_c^{(2)} + \delta_c^{(3)} + \delta_c^{(X)} \, .
\]
Typically the volume of the field subject to clipping is less than $1\%$, as seen in Figure \ref{fig:millClip}. The clipped field $\delta_c(x)$ initially possesses a nonzero mean. We subtract this value to restore the condition $\langle \delta(x) \rangle = 0$, however this has no impact on the power spectrum for $k>0$. 

Central to this work are the following three assertions, which we shall subsequently explore in sequence:

\begin{itemize}
\item{When subject to clipping, a GRF exhibits close to a \emph{scale-independent} suppression of its spectral power provided the logarithmic slope is not too steep $\left(|n| \lesssim 3 \right)$, and the clipping remains weak $\left(\delta_0 \gtrsim \sigma \right)$.  } 

\item{Higher powers of Gaussian fields $\delta_G^n$ experience a substantially stronger suppression of power at larger values of $n$. This is due to the extended tails of their probability density functions which are the dominant source of their variance.}

\item{In a cosmological context, the above implies that the power spectrum associated with the clipped density field may be well described by a truncated series expansion with appropriate suppression factors

\[
\eqalign{ \label{eq:PkClip}
P_{c}(k) &\equiv \langle  | \delta_c(k) |^2 \rangle \, , \cr
            &=  \sum_{i,j=1,2,3,X} A_{ij} \langle \delta_c^{(i)}(k) \delta_c^{*(j)}(k) \rangle  \, , \cr
            &\simeq  A_{11} P_L(k) + A_{13} P_{13}(k) + A_{22} P_{22}(k)  \, ,
}
\]
which follows from (\ref{eq:components}), since the strong $P_{XX}(k)$ and $P_{Xi}(k)$ terms generated by $\delta^{X}$ in (\ref{eq:expansion}) are now negligible due to the small values of $A_{XX}$ and $A_{Xi}$. The one-loop contributions to the power spectrum, $P_{22}(k)$ and $P_{13}(k)$, may be expressed in terms of the linear power spectrum $P_L(k)$, see e.g. \cite{BernardeauScoccimarro2002}.
}
\end{itemize}

\section{Spectral Response} \label{sec:spectrum}

The power spectrum of a Gaussian Random Field which has been subject to the clipping transformation defined by (\ref{eq:clip}) may be expressed in terms of the original power spectrum $P(k)$ and the threshold value $\delta_0$ as follows

\[ \label{eq:clipAnalytic}
\eqalign{
 P_c(k) =   \frac{1}{4} & \left[ 1 + \mathrm{erf}\left( \frac{\delta_0}{\sqrt{2} \sigma} \right) \right]^2  P(k) \, + \cr
 & \sigma^2 \sum_{n=1}^{\infty} C_{n} \left( \frac{\delta_0}{\sqrt{2} \sigma} \right) \hat{P}^{*(n+1)}(k) \, ,
 }
\]
where  $\sigma$ is the standard deviation of the field prior to clipping, $\hat{P}(k)$ denotes the input power spectrum normalised such that the corresponding correlation function has an amplitude of unity at zero lag $(\hat{P}(k) = P(k)/\sigma^2)$, $\hat{P}^{*n}(k)$ represents a self-convolution of the order $n$ [for example $\hat{P}^{*2}(k) = \hat{P}(k)*\hat{P}(k)$], and $C_n(x)$ is the distortion coefficient 

\[ \label{eq:cn}
C_n(x) = \frac{H_{n-1}^{2} \! \left( x \right)}{\pi 2^n (n+1)!}  e^{-2x^2} \, ,
\]
where $H_n$ is a Hermite polynomial. Further details of this calculation may be found in Appendix \ref{sec:app}, along with a similar expression for higher powers of GRFs. The first term in (\ref{eq:clipAnalytic}) is a scale-independent suppression of the original power, while the second quantifies the form of spectral distortion which occurs. From this model it is apparent that an input signal with a flat spectrum $(n=0)$ will remain flat. Furthermore, when $\delta_0>\sigma$, inputs with only modest spectral slopes will experience little distortion, since the first term in (\ref{eq:clipAnalytic})  dominates, while the contribution from the second term will not greatly differ in its shape from the first. 

The above provides us with a useful starting point, although our analysis will differ slightly since we shall be working with evolved cosmological perturbation fields which are non-Gaussian.

\section{Amplitude Coefficients} \label{sec:coeffs}

In order to take full advantage of the prescription given by (\ref{eq:PkClip}) it is important to know the values of the amplitude coefficients $A_{ij}$ for a clipped field. In this section we shall explore methods for estimating these values, which are verified by clipping random realisations of three-dimensional fields. 

\subsection{Analytic Models}

To predict the response of a GRF to clipping, the linear amplitude parameter $A_{11}$ as defined in (\ref{eq:PkClip}) is readily evaluated by considering the fractional change in variance of the probability density function $p(\delta)$. As we saw in (\ref{eq:clipAnalytic}), provided the clipping is weak and the spectral slope is not too steep, negligible spectral distortion occurs.  Since the shape of $P(k)$ is well preserved then the reduction in power must match the fractional change in the correlation function at zero lag, $\xi(0)$.  This is given by

\[ \label{eq:A1clip}
\eqalign{
{A}_{11}(\delta_0) &= \frac{ \mathrm{var} \left[ \delta_c(x)\right]}{ \mathrm{var} \left[ \delta(x)  \right]} \cr
&=  \frac{\int_{-1}^{\dThresh} p(\delta) (\delta - \bar{\delta}_c)^2 \ud \delta + \int_{\dThresh}^{\infty} p(\delta)(\dThresh - \bar{\delta}_c)^2 \ud \delta}{ \int_{-1}^{\infty} p(\delta) \delta^2 \ud \delta  } \, ,
}
\]
where  $\dThresh$ is the threshold at which clipping occurs, as defined in (\ref{eq:clip}). The mean value of the clipped field is $\bar{\delta}_c$, and $p(\delta)$ is the probability density function of the original GRF. As expected, ${A}_{11}$ approaches unity for very large values of $\delta_0$, and vanishes when $\delta_0=-1$. 

Next we consider a hybrid field comprising a linear combination of a GRF and its square 

\[ \label{eq:alphabeta}
\delta_t = \alpha \delta_G + \beta \left( \delta_G^2  - \langle \delta_G^2 \rangle \right) \,  .
\]
The power spectrum associated with the total field $\delta_t$ is simply given as a weighted sum of the two constituent spectra $P_t(k) = \alpha^2 P_L(k) + \beta^2 P_{22}(k)$. Once clipping is applied, both contributions are suppressed. But clipping also violates the symmetry of the field, therefore the covariance of the two component fields $ \langle \delta_{Gc} \delta_{Gc}^2 \rangle$ no longer vanishes. Consequentially, we introduce the cross-spectrum term $ P_{12}(k)$ 

\[ \label{eq:clipPower}
P_c(k) \simeq \alpha^2  A_{11} P_L(k) + \beta^2 A_{22} P_{22}(k)  +  \alpha \beta P_{12}(k) \, ,
\]
where the second-order amplitude coefficient $A_{22}$ follows a similar prescription to $(\ref{eq:A1clip})$, except now the functional form of $p(\delta)$ is that of a squared Gaussian, suitably compensated to ensure a mean of zero. 

\[ \label{eq:A2clip}
\eqalign{
{A}_{22}(\delta_0) &=\frac{ \mathrm{var} \left[ \delta^2_{Gc}(x)\right]}{ \mathrm{var} \left[ \delta^2_G(x)  \right]}  \, .
}
\]
The threshold experienced by the second-order field $\delta^2_{Gc}$ is not $\delta_0$ but is determined by solving (\ref{eq:alphabeta}) to find the value of $\delta_G$ which satisfies the condition $\delta_t = \delta_0$. This provides the upper and lower bounds on the underlying field $\delta_G$, beyond which the hybrid field  $\delta_t$  is subject to clipping. The cross-spectrum $P_{12}(k)$ may be evaluated by considering the covariance of the two fields

\[
P_{12}(k) =  \frac{2 \, \mathrm{cov} \left( \delta^2_{Gc}, \delta_{Gc} \right)}{\sqrt{\mathrm{var} (\delta^2_{Gc})  \mathrm{var}  (\delta_{Gc}) }}\sqrt{A_{11} A_{22} P_{11}(k)   P_{22}(k)} \, ,
\]
however we find that if the clipping threshold is high then this term is negligible.

Extending the test field defined in (\ref{eq:alphabeta}) to include a third power $\delta_G^3$ leads to an additional $A_{13} P_{13}(k)$ and $P_{23}(k)$ terms when clipped. However this scenario is simplified by the equivalence of the $A_{22}$ and $A_{13}$ terms

\[ \label{eq:A22A13}
\eqalign{
A_{22} &= \frac{\langle \delta_c^2 \delta_c^2 \rangle}{\langle  \delta_G^2 \delta_G^2 \rangle} \cr
&= \frac{\langle  \delta_c \delta_c^3 \rangle}{\langle  \delta_G \delta_G^3 \rangle} \cr
&= A_{13} \, .
}
\]
Thus we find that, provided the clipping is weak, the following simple model provides a good estimate of the clipped power spectrum
\[ \label{eq:clipModel}
P_c(k) \simeq   A_{11} P_L(k) +  A_{22} \left[ P_{22}(k)  +  P_{13}(k) \right]  \, ,
\]
which we shall make use of in \S \ref{sec:results}.

Finally we consider the implications of introducing an unknown component, the $\deltaX$ term given in (\ref{eq:expansion}). For clipping to be effective we require $\deltaX$ to be subdominant in the regions of moderate density where $\delta < \delta_0$. If this condition is satisfied, and the remainder of the field is of a known functional form, it is relatively straightforward to predict the clipped power spectrum.  Since the nature of $\deltaX$ is unknown it is necessary to explore a range of $\delta_0$ values in order to reveal the point at which the contributions from $\deltaX$ become negligible. We should be able to extract consistent results from a range of different $\delta_0$ values provided they are strong enough to remove $\deltaX$, and yet weak enough to maintain the validity of our model  (\ref{eq:clipModel}).  Another useful consistency check is to repeat the analysis using a field defined by a different grid size or smoothing length. If the smoothing length is too fine, the fluctuations will be larger and it may prove more difficult to model the clipped field. Conversely, choosing too coarse a smoothing length leads to inefficient clipping, as a greater volume must be subject to clipping in order to achieve the desired result. A larger smoothing length also reduces the range of available $k$ modes. While our fiducial selection of a $256^3$ grid corresponds to a Nyquist frequency of $k_N \simeq 1.6 \hmpc$, the windowing correction close to the Nyquist frequency is unreliable, so we conservatively impose a maximum wavenumber of $k_{\mathrm{max}}=0.7 \hmpc$. 

\subsection{Numerical Results}

Before working with cosmological density fields, we first consider the response of a GRF with a power-law spectrum, $P(k) \propto k^n$, when subject to clipping as defined by (\ref{eq:clip}). The simplest case is white noise, $n=0$. When subject to clipping, white noise remains white, experiencing a purely scale-independent suppression of power. However for more general power spectra, some spectral distortion does arise, as given by the convolution term in (\ref{eq:clipAnalytic}). The level of distortion increases with the severity of the clipping and the slope of the power spectrum. Since the threshold values used in this work are weak, typically less than $1\%$ of the volume experiences clipping, we do not find it necessary to include this distortion, though it may be desirable for higher-precision studies.   Even when neglecting the spectral distortion, the prescription given by (\ref{eq:clipPower}) recovers the clipped power $P_c(k)$ to better than $1\%$, provided $A_{11}>0.5$.

\begin{figure}
\includegraphics[width=80mm]{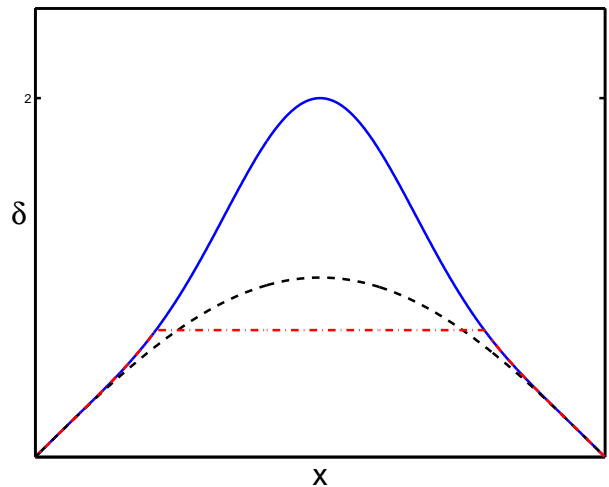}
\caption{Simplified schematic of how a nonlinear field (solid) and linear field (dashed) both reduce to a similar form when subject to clipping (dot-dashed).  \label{fig:schematic}}
\end{figure}

\begin{figure}
\includegraphics[width=80mm]{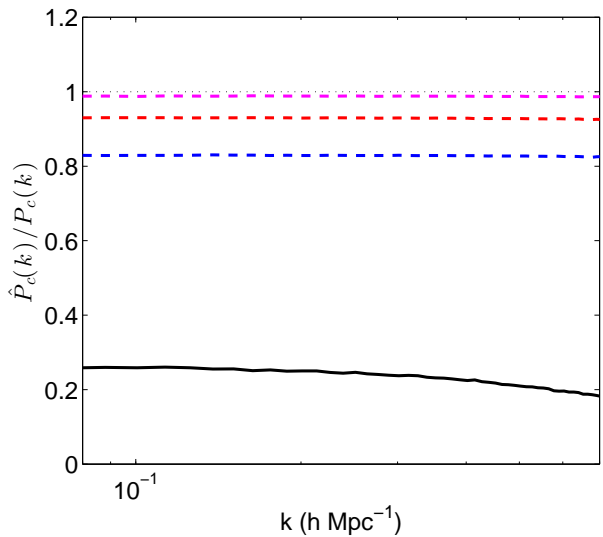}
\caption{A demonstration of the removal of higher order contributions to the power spectrum using clipping. The reference field is the density perturbation from the $z=127$ snapshot of the Millennium-I simulation, amplified in accordance with linear theory such that its standard deviation $\sigma(\delta_G) = 0.5$. The solid line demonstrates the large underestimation of the power spectrum due to the presence of an extra component $\delta^{X} = \delta^5$. The dashed lines illustrate the rapid improvement in the theoretical model after applying clipping thresholds at the $2$, $1.5$ and $1 \sigma$ levels. Clipping at $1 \sigma$ recovers the power spectrum to within $1\%$.  \label{fig:fake}}
\end{figure}

For illustrative purposes we consider a very simple model for a non-linear field by introducing an ``unknown" component, acting as the $\delta^{(X)}$ term from (\ref{eq:expansion}). The total field is chosen to be

\[ \label{eq:alphabetagamma}
\delta_t =  \delta_G   +  \delta_G^5 \, .
\]
The selection of $\delta_G^5$ is somewhat arbitrary, although it is relevant to the later sections where it generates one of the leading terms neglected by our theoretical model (one-loop perturbation theory) due to its covariance $\langle \delta_G  \delta_G^5 \rangle$ leading to $P_{15}(k)$. For our reference GRF we use the $z=127$ dark matter field from the Millennium-I simulation \cite{2005Natur.435..629S}. The perturbation amplitudes are scaled in accordance with linear theory such that their standard deviation $\sigma(\delta_G) = 0.5$. Figure \ref{fig:schematic} provides a simple schematic of this field, and how clipping eliminates most of the excess contribution from the $\delta_G^5$ term. In Figure \ref{fig:fake} the solid line illustrates the large underestimation of the power spectrum due to the presence of the extra component $\delta^{X} = \delta_G^5$. The dashed lines illustrate the substantial improvement in the theoretical model after applying clipping thresholds of $2 \sigma$ (top), $1.5 \sigma$ (middle) and $1 \sigma$ (bottom), where $\sigma$ refers to the standard deviation of $\delta_G$. The $1 \sigma$  threshold $(\delta_0 = 0.5)$ reduces the discrepancy between the true and theoretical power spectra to less than $1\%$.

\section{Cosmological Fields} \label{sec:cosmo}

As in \citet{SimpsonClip}, we make use of data from the Millennium-I simulation  \cite{2005Natur.435..629S}, and adopt the same resolution  ($256^3$; box size $= 500 \hinvmpc$). We begin with a study of how the dark matter power spectrum responds to clipping, before extending our analysis to consider a mock galaxy catalogue from \citet{Guo2011}.  For each field under consideration we explore a range of threshold values $\delta_0$. If the threshold is too high, insufficient nonlinear structure has been removed from the field and the power spectrum cannot be well described by (\ref{eq:PkClip}) at larger values of $k$. If the threshold is too low, then the spectral distortion in (\ref{eq:clipAnalytic}) becomes large, and for galaxy fields their shot noise will become more problematic due to the strongly suppressed amplitude of the power spectrum. In addition to a goodness-of-fit test, the bispectrum results from \citet{SimpsonClip} provide a useful indicator for the range of $\delta_0$ and $k$ values over which we can expect our model to be valid. 

\subsection{Dark Matter Fields}

At very high redshifts where the field is linear, we can utilise the analytic solution for $A_{11}$ given by (\ref{eq:A1clip}). At lower redshifts, the one-loop power spectrum is almost an order of magnitude greater than the linear power spectrum as we approach $k \sim 1 \hmpc$. The ratio $A_{11}/A_{22}$ is not great enough to overcome this, so we must take into account the presence of the one-loop term if we are to utilise the clipped power spectrum out to higher frequencies. This poses a significant challenge, since the higher order terms $\delta^{(2)}$ and $\delta^{(3)}$ are not simple powers of the underlying field. For example \cite{1995Catelan}

\[ \label{eq:delta2}
\delta^{(2)} = \frac{5}{7} \delta^{(1)2} - a \left( u^{(1)} \cdot \nabla  \right)  \delta^{(1)} + \frac{2}{7} a^2 \sum_{\alpha \beta} \left( \partial_\alpha u^{(1) \beta} \right)^2 \, ,
\]
where $a$ is the scale factor and $u^{(1)}$ represents the linear velocity field. Clearly there is no local solution for $\delta^{(2)}$ and $\delta^{(3)}$ given $\delta^{(1)}$. To proceed we form an approximate model for the amplitude coefficients by introducing a single free parameter $x_i$ which relates the fraction of mass removed from the field $f_m$ with the fraction of mass removed from the component field $\delta^{(i)}$

\[ \label{eq:fmass}
f_i = x_i f_m \, .
\]
The two fractions $f_1$ and $f_2$ may then be used to generate $A_{11}(f_m)$ and $A_{22}(f_m)$ respectively, following (\ref{eq:A1clip}) and (\ref{eq:A2clip}). In effect this assumes that the functional form of $p(\delta^{(2)})$ matches that of  a squared GRF, $p(\delta^{(1)2})$.

\subsection{Galaxy Fields}

\begin{figure}
\includegraphics[width=80mm]{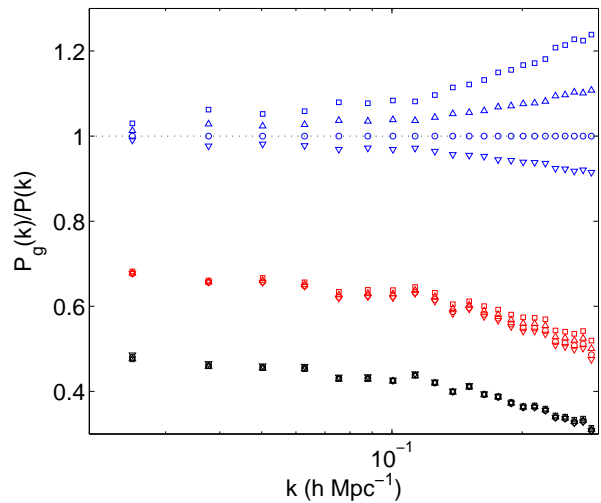}
\caption{This illustrates the suppression of nonlinear galaxy bias when clipping is applied. The upper set of four points corresponds to the fractional change in the power spectra  induced by four different values of $b_2 = [-0.01; 0; 0.01; 0.02]$, relative to the fiducial $b_2=0$ power spectrum. The middle and lower sets of points shows the same range of values for $b_2$, but now having clipped $5\%$ and  $10\%$ of the mass respectively. The fractional change induced in the power spectrum by the presence of a non-zero $b_2$ term is reduced by more than an order of magnitude after clipping is applied. The linear bias is fixed at $b_1 = 0.9$, and the grid size is $7.8 \hinvmpc$.   \label{fig:b2}}
\end{figure}

In the previous subsection we assumed that the observed field is the matter density, and that the cosmological parameters are  known. Here we extend our analysis to the more practical case of a galaxy field. A common assumption in cosmological studies is that on large scales any perturbation in the number density of galaxies is directly proportional to the underlying density perturbation. This is the case of local linear bias, $\delta_g = b \delta_m$.  More generally, local deterministic models of galaxy bias may be represented as a power series \cite{fry1993biasing}

\[
\delta_g = \sum b_i \frac{\delta_m^k}{k!} \, .
\]
As with the power series expansion associated with the dark matter (\ref{eq:expansion}), higher-order terms are subject to significantly stronger suppression by clipping than the leading order term. Therefore a clipped galaxy number density field will more closely adhere to the linear bias prescription than the original field. This is illustrated in Figure \ref{fig:b2}, where the impact of the $b_2$ parameter on the form of a mock galaxy power spectrum is found to diminish by more than an order of magnitude after clipping is applied. For a more detailed calculation of this linearisation process, see \S \ref{sec:clipBias}. 

When we determine the best-fit value of an amplitude parameter $A_{ij}$ at a given redshift slice, we do so without knowledge of the galaxy bias or the true amplitude of the dark matter field. Therefore what we are actually measuring is 

\[ \label{eq:A1prime}
A'_{11} = b^2 \left[ \frac{\sigma_8(z)}{\, \sigma_8^{\rm{fid}}(z)} \right]^2 A_{11}  \, ,
\]

\noindent and more generally 

\[ \label{eq:Aijprime}
A'_{ij} =  b^2\left[\frac{\sigma_8(z)}{\sigma_8^{\rm{fid}}(z)}  \right]^{i+j} A_{ij} \, ,
\]
while the mass fraction $f_m$ is also rescaled

\[ \label{eq:rescaleF}
f_m' = b \frac{\sigma_8(z)}{\sigma_8^{\rm{fid}}(z)}  f_m \, .
\]

\noindent Ordinarily, when we can only make use of the linear power spectrum on very large scales $k<0.1 \hmpc$, there exists a perfect degeneracy between the values of $b$ and $\sigma_8$, as is apparent from (\ref{eq:A1prime}). With clipping, not only are we able to extend our analysis to higher $k$ modes due to the suppression of the highly nonlinear terms, but we also lift the degeneracy between the bias and $\sigma_8$ parameters, due to these higher-order terms.   We shall therefore seek to measure simultaneously the linear bias parameter $b$ and $\sigma_8$ from the clipped galaxy power spectrum. 

\begin{figure*}
\includegraphics[width=170mm]{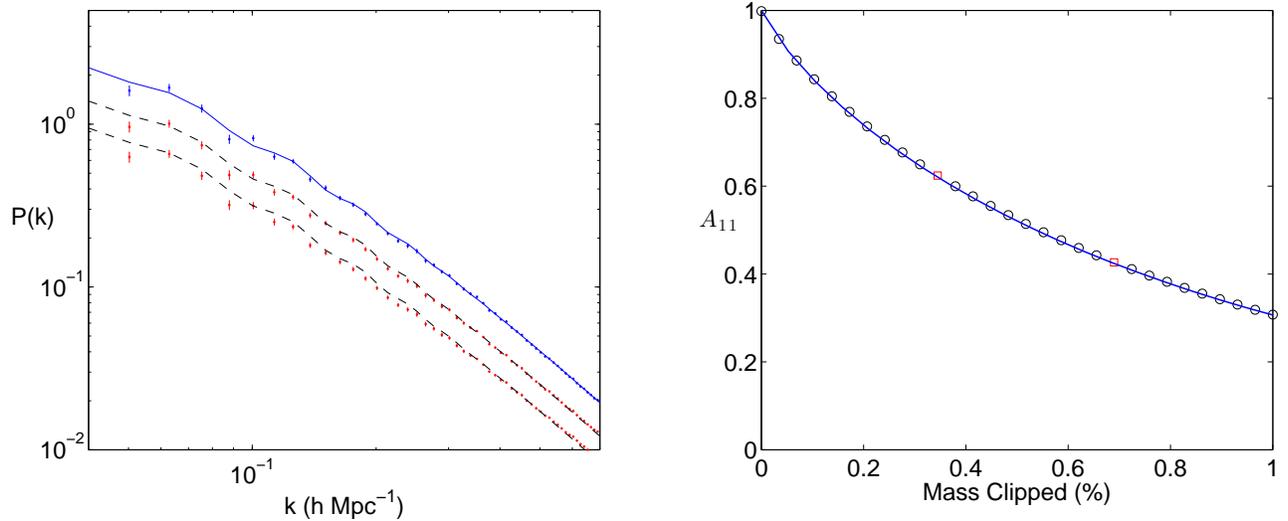}
\caption{ \label{fig:clip127}    Reduction in the power spectrum of density perturbations from the $z=127$ snapshot of the Millennium-I simulation, subject to clipping as defined in (\ref{eq:clip}). \emph{Left:} The upper set of data points correspond to the case of the unclipped field, while the middle and lower sets of points are the power spectra obtained when  removing 0.3 per cent and 0.7 per cent  of the mass respectively. The solid line is the linear theory power spectrum from CAMB, while the dashed lines are evaluated by rescaling the amplitude of the solid line in accordance with (\ref{eq:A1clip}).   \emph{Right:} The solid line represents the analytic prescription given by  (\ref{eq:A1clip}), as a function of the fraction of mass removed by clipping, while the circular points denote the best-fit amplitude determined by fitting the theoretical power spectrum to the data. The two square markers in the right hand panel correspond to the two clipping fractions illustrated in the left panel. }
\end{figure*}

\begin{figure*}
\includegraphics[width=170mm]{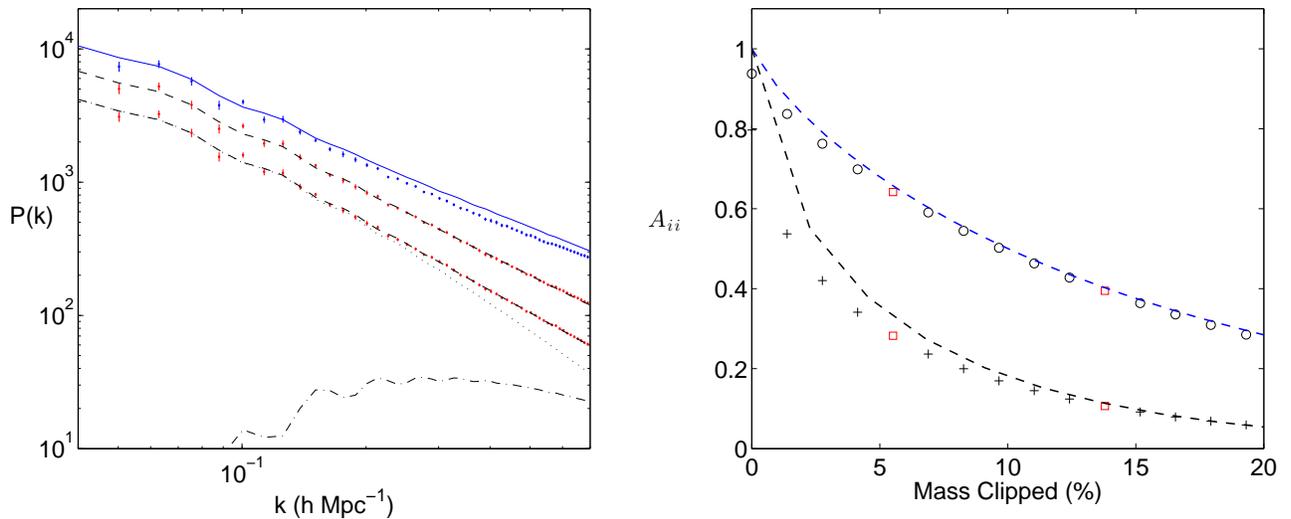}
\caption{ \label{fig:clipz0687}    Similar to Figure \ref{fig:clip127} except here we are using the dark matter field at z=0.687. We therefore need to introduce the one-loop contributions to the power spectrum. \emph{Left:} The solid line is the power spectrum predicted by one-loop perturbation theory, which can be seen to overestimate the data points on small scales. The dashed lines are a combination of suppressed linear and one loop power spectra, fitted with two degrees of freedom, $A_{11}$ and $A_{22}$. The upper and lower dashed lines correspond to clipping fractions of 6 and 14 per cent (by mass), and for the latter the theoretical model is decomposed into the linear (dotted) and one-loop (dot-dash) components. \emph{Right:}  The circles and crosses are the values of $A_{11}$ and $A_{22}$ obtained by simultaneously fitting the linear and one-loop spectra to the clipped power spectra over the range $0.1 < k < 0.7 \hmpc$.  The dashed lines represent the theoretical estimate for the functions $A_{11}$ and $A_{22}$, derived from (\ref{eq:A1clip}), (\ref{eq:A2clip}), and (\ref{eq:fmass}). Each line has one degree of freedom to model the unknown fraction of the mass removed from each component field relative to the total field. Note that at low values of $f_m$, below $5\%$, we should not expect the data to match the theory, since insufficient clipping has occurred. }
\end{figure*}

\begin{figure*}
\includegraphics[width=170mm]{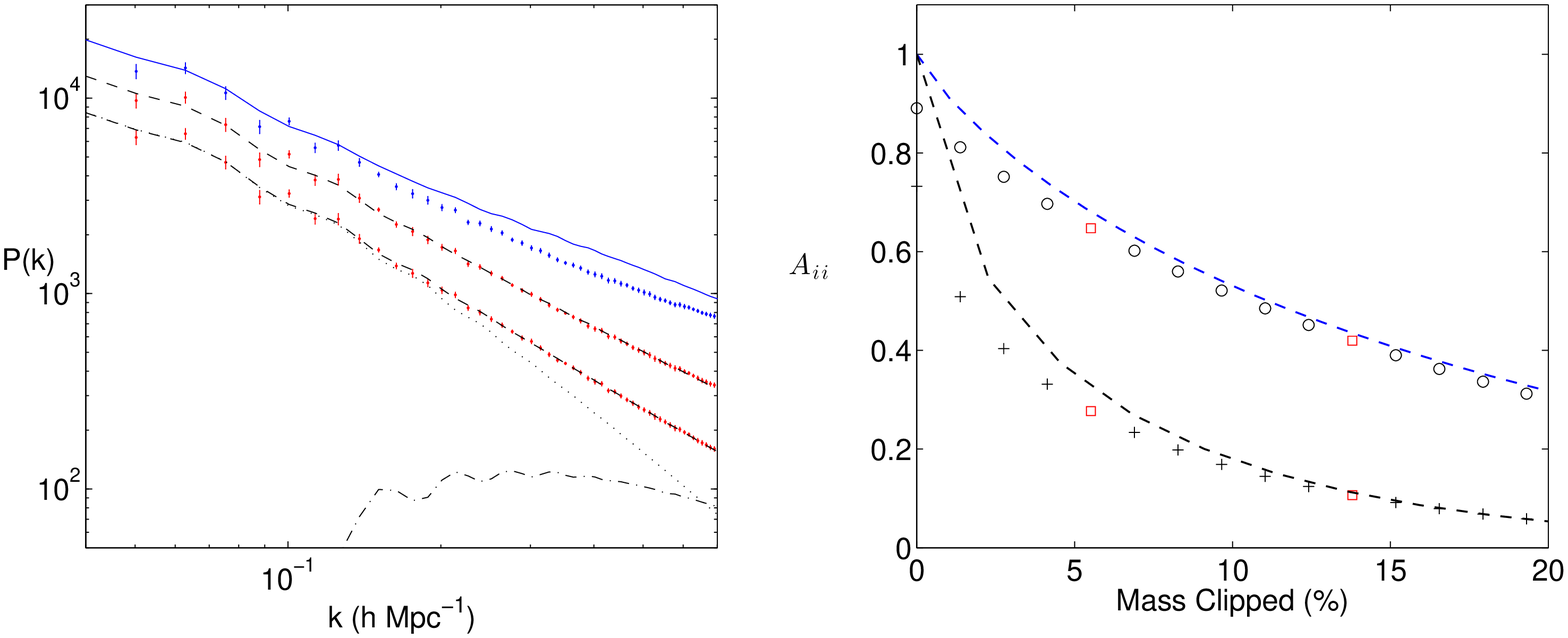}
\caption{ \label{fig:clipz0}     Same as Figure \ref{fig:clipz0687} except here we are using the dark matter field at z=0. }
\end{figure*}

\begin{figure*}
\includegraphics[width=170mm]{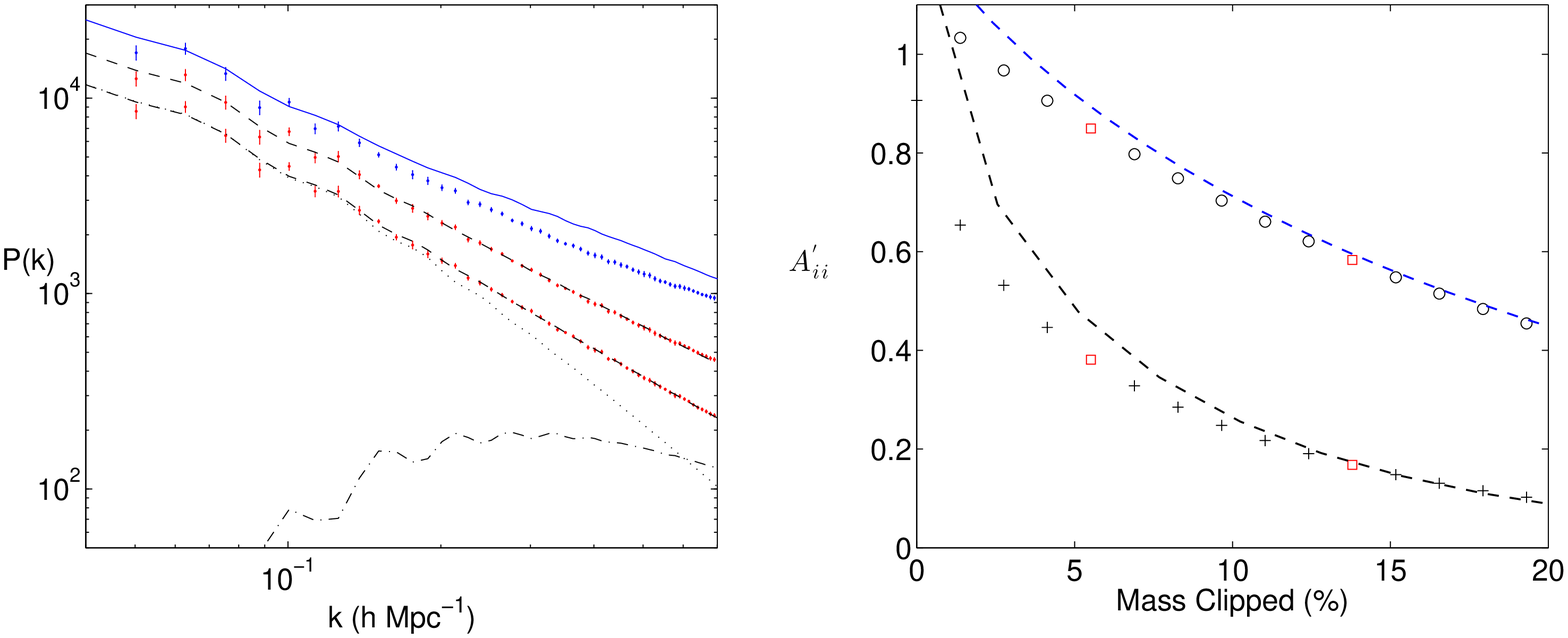}
\caption{ \label{fig:clipGal}    Similar format to Figure \ref{fig:clipz0687} except here we are using the a mock galaxy catalogue at $z=0$. \emph{Left:}  The galaxy power spectrum responds to clipping in a very similar manner to the dark matter power spectrum in Figure \ref{fig:clipz0}. \emph{Right:} The dashed lines are evaluated by rescaling those from the right hand panel of Figure \ref{fig:clipz0}, using (\ref{eq:Aijprime}) and (\ref{eq:rescaleF}) to account for the galaxy bias.}
\end{figure*}

\begin{figure}
\includegraphics[width=80mm]{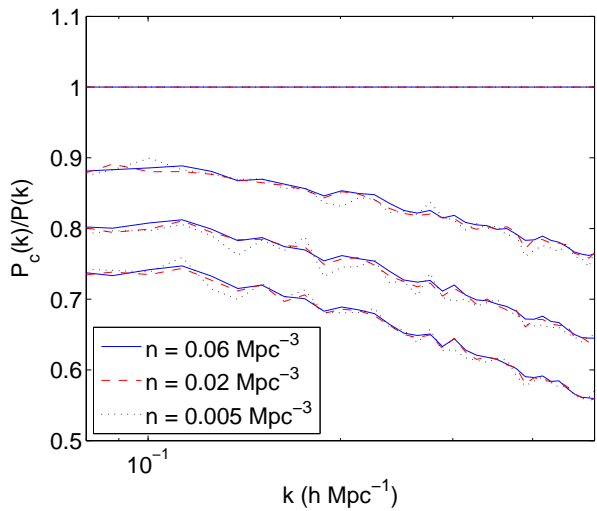}
\caption{Ratio of the clipped galaxy power spectrum to the original power spectrum, for three different number densities (solid, dashed, and dotted) and three different clipping thresholds. These appear to have a negligible impact on the clipped power spectra. Lower number densities cannot make use of the highest $k$ values, due to the greater shot noise \label{fig:nbar}}
\end{figure}

\begin{figure}
\includegraphics[width=80mm]{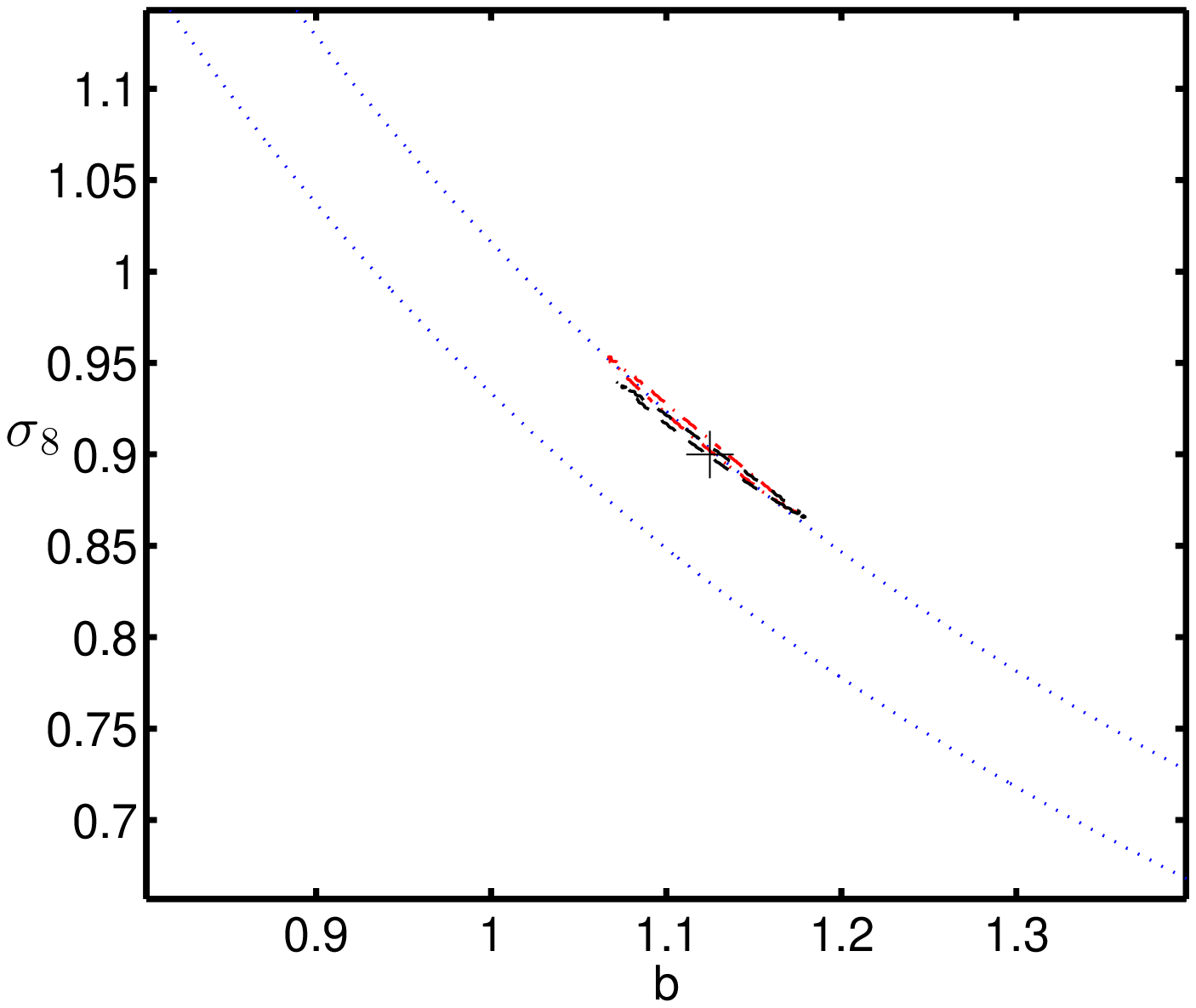}
\caption{The dashed and dot-dashed contours designate 95\% confidence limits, generated by clipping $5\%$ and $10\%$ of the galaxies respectively. The black cross at $(b=1.125, \sigma_8 = 0.9)$ designates the true values of $\sigma_8$ and the linear bias as determined by the ratio of the galaxy and dark matter power spectra on the largest scales.  The dotted lines represent the confidence contours without clipping, which leaves us restricted to the linear regime. }
\label{fig:massContours}
\end{figure}

\section{Results}
\label{sec:results}

The techniques developed above are now applied to dark matter fields at three different redshifts, and to a mock galaxy field at $z=0$. The model we use for the clipped power spectrum is given by

\[ \label{eq:clipModelPrime}
\hat{P}_c(k) =   A'_{11} P_L(k) +  A'_{22} \left[ P_{22}(k)  +  P_{13}(k) \right]  \, .
\]

\subsection{Dark Matter}

At high redshift the dark matter density is known to be very well represented by a GRF \cite{Planck_fNL}, and we can therefore model its response to clipping using the prescription given by (\ref{eq:clipModelPrime}) and (\ref{eq:A1clip}). The left hand panel of Figure \ref{fig:clip127} illustrates how clipping the cosmological density field using the $z = 127$ snapshot leads to a largely scale-independent drop in power.  The spectral distortion induces a significant discrepancy at the lowest $k$ modes,  therefore we exclude the largest modes $\left( k<0.1 \hmpc \right)$, from consideration throughout this work.  The error bars are established by considering the variance of the power spectra derived from each octant of the cube. The right panel plots the linear amplitude parameter $A_{11}$ as a function of the fractional mass removed by the clipping process. The data points arise from fitting the theoretical linear power spectrum from CAMB \cite{2000CAMB} to the clipped data, while the functional form of the solid line is given by (\ref{eq:A1clip}).

Moving to lower redshifts inevitably introduces greater complexity, and the need to introduce the one-loop power spectrum. Ordinarily the one-loop term is considerably larger than the linear power spectrum at $k \sim 0.5 \hmpc$ for evolved fields. Therefore even when suppressed by clipping, the one-loop term is of comparable magnitude to the linear contribution. We model the clipped power spectrum in accordance with (\ref{eq:clipModelPrime}). Equating $A_{13}$ and $A_{22}$ arises from approximating $\delta^{(2)}  \propto \delta^2$ and $\delta^{(3)}  \propto \delta^3$. We continue to treat the errors associated with $P(k)$ as uncorrelated which ordinarily would be a very poor description, but as we shall see later, the covariance of Fourier modes is greatly reduced by clipping, as the influence of nonlinear structure is significantly reduced. 

We can readily extend this to include the two-loop components, however for simplicity we restrict our analysis to the above model. Similarly we do not compensate for spectral distortions or introduce the $P_{12}(k) $ term. Instead we restrict ourselves to large threshold values, where these effects remain small. We define this weak clipping regime to be $A_{11}>0.5$, so the amplitude of the linear power spectrum is suppressed by less than a factor of two. It is in any case desirable to make use of reasonably large values of $A_{11}$ in order to maintain a strong signal. We also note that for more evolved fields a larger proportion of the total mass needs to be removed in order to achieve an equivalent value of $A_{11}$.

In Figure \ref{fig:clipz0687} we can see the influence of various clipping fractions applied to the dark matter field at $z=0.687$. In the left panel, the solid line represents the theoretical one-loop power spectrum while the blue dots represent the unclipped datapoints.  In the absence of clipping, the theoretical model breaks down at approximately $k \sim 0.1 \hmpc$.  The two lower sets of points and dashed lines correspond to different clipping fractions. The dashed and dot-dashed lines illustrate the two components of the lower clipped spectrum, the linear and one-loop power spectra respectively.  As before, the circles in the right hand panel of Figure \ref{fig:clipz0687} illustrate the best fit amplitudes for the linear power spectrum, $A_{11}$, as a function of clipping fraction. The red markers denote the two clipping fractions used in the left hand panel. The crosses denote the best-fit values for $A_{22}$, which controls the amplitude of the one-loop power spectrum. We note that the goodness of fit is considerably improved once more than $5\%$ of the mass has been removed. 
The dashed lines illustrate the behaviour we ought to expect if the field was simply of the form (\ref{eq:alphabeta}). We have used the approximation (\ref{eq:fmass}), so each dashed line has one degree of freedom. Note that we should expect there to be a strong disagreement between the theoretical curves and data points at low clipping fractions, since for $A_{11}>0.75$ the clipping is too weak to allow one-loop perturbation theory to adequately model the power spectrum across the full range of wavenumbers used in the fit $(0.1 \hmpc < k < 0.7 \hmpc)$. The discrepancy in $A_{22}$ for $f_m > 5\%$ appears approximately consistent with the expected amplitude of the two-loop term, which falls with $A_{33}$. 
We shall leave a more thorough investigation of the amplitude coefficients and their cosmological dependence for future work.  Instead we calibrate their values from the dark matter field, which we subsequently apply to a mock galaxy population.

Moving to the $z=0$ snapshot, where the linear perturbation amplitude is $\sigma_8=0.9$, the prominence of nonlinearities is further enhanced, as is the departure from perturbation theory in the unclipped field. Figure \ref{fig:clipz0} is in the same format as Figure \ref{fig:clipz0687}, and we see a very similar response to the clipping process. In the left hand panel, the contribution from the one loop power now exceeds that of the linear power spectrum at $k=0.7$. In the right hand panel the fractional suppression of the one-loop term, $A_{22}$, matches that from the $z=0.687$ field to better than $0.2\%$ for larger clipping fractions, despite the one-loop power growing by a factor of four between these redshifts. 

All of the analyses above were repeated using a coarser smoothing length $(128^3; \sim 3.9 \hinvmpc)$, and this was found to generate consistent results.

 \begin{figure*}
\includegraphics[width=180mm]{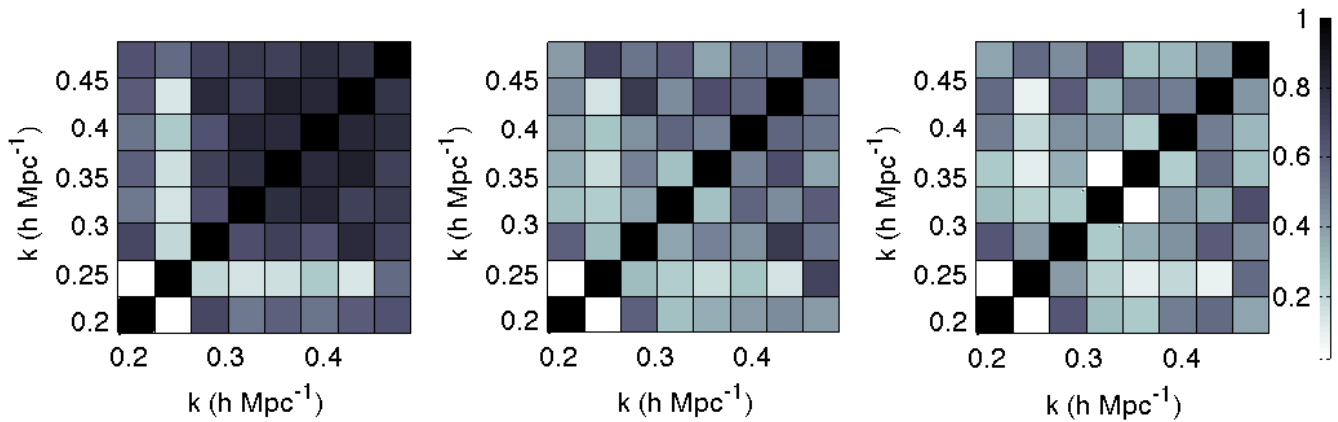}
\caption{ \label{fig:pk_corr_panels} Decorrelation of the galaxy power spectrum. \textit{Left:} The left panel shows the correlation matrix for the power spectrum derived from a mock galaxy catalogue in the Millenium-I simulation at $z=0$, with a mass cut $\log_{10}(M_\ast/M_\odot h) \ge 9$.  \textit{Middle} The same as the left-hand panel, but now using a field to which clipping has been applied to $0.1\%$ of the volume.  \textit{Right:}  The same as the central panel, but now clipping $1\%$ of the volume.}
\end{figure*}

\subsection{Galaxies} \label{sec:galaxies}

In Figure \ref{fig:clipGal}  we see the clipped galaxy field responds with very similar behaviour to that of the dark matter field. In the right hand panel the  dashed lines are identical to those in Figure \ref{fig:clipz0}, except we have scaled their values in accordance with equations  (\ref{eq:Aijprime}) and (\ref{eq:rescaleF}), accounting for the linear bias $b = 1.125$ which we estimate from the ratio of the unclipped galaxy and dark matter power spectra on large scales. 

Since we wish to explore P(k) at high values of k where P(k) is relatively low, and since clipping suppresses the amplitude further, it is desirable to use a galaxy catalogue with a relatively high number density. In Figure \ref{fig:nbar} we illustrate the change in the galaxy power spectrum with three different number densities, each subject to three different threshold values. There is no evidence to suggest the resulting clipped power spectrum is sensitive to the number density of galaxies, although if the smoothing length used to determine the field $\delta_g$ becomes too small then this will inevitably become problematic. For all the examples we have been considering at lower redshifts, the fractional volume subject to clipping $f_{\rm{v}}$ is very small, typically less than one per cent, so the reduction to the shot noise is negligible. 

For our likelihood analysis, we fix all cosmological parameters except for the galaxy bias $b$ and the amplitude of matter perturbations $\sigma_8$. To arrive at the confidence contours shown in Figure \ref{fig:massContours}, we follow the following procedure:

\begin{itemize}
\item{Evaluate the power spectra associated with the $z=0$ dark matter field for a range of clipping thresholds. At each, we perform a  simultaneous best-fit for the two amplitude parameters given in (\ref{eq:clipModelPrime}). We use CAMB  \cite{2000CAMB} and  CPT \cite{CPTtaruya} codes to generate the theoretical $P_L(k)$, $P_{13}(k)$, and $P_{22}(k)$ terms. This results in the data points illustrated in the right hand panel of Figure \ref{fig:clipz0}. These points are interpolated to form the functions $A_{11}(f_m)$ and $A_{22}(f_m)$. }
\item{For a given set of values for $b$ and $\sigma_8$, we use the relations (\ref{eq:Aijprime}) and (\ref{eq:rescaleF}) to establish the $A_{ij}'(f_m)$ parameters we expect to observe under these conditions.}
\item{The galaxy field is constructed in the same manner as the dark matter field, and the desired threshold is applied before evaluating the galaxy power spectrum.}
\item{
For the galaxy power spectrum we correct for the shot noise by subtracting $P_s(k) = 1/{\bar{n}}$ from the observed power, where $\bar{n}$ is the number density in the unclipped field. Clipping will act to reduce the magnitude of $P_s(k)$, but since the volume fraction $f_v$ is typically less than $1\%$, this correction is negligible.}
\item{As with the dark matter, errors are estimated by subdividing the simulated cube and evaluating the sample variance. As expected, returning the galaxy field closer to a Gaussian state leads to the decorrelation of neighbouring Fourier modes. This is illustrated in Figure \ref{fig:pk_corr_panels}, where the application of clipping substantially reduces the strength of the off-diagonal terms in the correlation matrix. We assume a diagonal covariance matrix. }
\item{Using the prescription of (\ref{eq:clipModelPrime}) we evaluate the predicted clipped field $\hat{P}_c(k)$ and compare this with the clipped galaxy field over the range ($0.1 \hmpc$ $< k <$ $0.7 {\hmpc}$), to evaluate the associated likelihood, and this procedure is repeated at each point in the plane of Figure \ref{fig:massContours}.}
\end{itemize}
The procedure outlined above leads to the dashed and dot-dashed contours of Figure \ref{fig:massContours}, when clipping $5\%$ and $10\%$ of the galaxies respectively. Clipping greatly increases the number of available modes, increasing the level of precision, while the one-loop term acts to lift the degeneracy between the linear bias and $\sigma_8$. Consistent results are found when using different clipping thresholds, and when using a coarser  smoothing length to define the fields. 

We emphasise that the above analysis takes place in real space, while any application to real data will require redshift space distortion effects to be taken into consideration. This is likely to reduce the efficiency of the clipping process, and may therefore lead to a reduction in the maximum accessible wavenumber. However since the location of the highest density regions remain identifiable, and since the galaxy bias becomes more linear as shown in Figure \ref{fig:b2}, we anticipate that substantial gains in the amount of  information extracted from galaxy redshift surveys will be achieved. 

\section{Conclusions} \label{sec:conclusions}
Conventional perturbation theory is limited in its success of modelling clustering statistics for two reasons. The addition of higher-order terms does not lead to a rapid convergence, meaning a high number of loops needs to be counted, which becomes increasingly complex at high order. It has proven difficult to push the range of applicability of perturbation theory beyond $\sim 0.2 \hmpc$ at the present epoch. This leaves a large range of wavenumbers $0.2 \lesssim k/(\hmpc) \lesssim 1$ where we understand the physics of the mass distribution well, but lack the mathematical tools, needing to rely on simulations. Secondly, perturbation theory relies on the velocity field being curl-free and single-valued, an assumption which breaks down for highly virialised structures. Both of these issues are addressed by the clipping process. The amplitude of higher-order terms decays much more rapidly when clipping is applied.  And it is the clipped regions, where the most massive halos lie, which most strongly violate the assumptions held when utilising perturbation theory. 

By modelling the clipped power spectrum as the sum of rescaled linear and one-loop power spectra (\ref{eq:clipModelPrime}), we have demonstrated that large wave numbers may be used ($k <  0.7 \hmpc$), greatly increasing the amount of information available. Furthermore a simultaneous measurement of the galaxy bias and $\sigma_8$ parameters can be performed without resorting to higher order statistics. 

The logarithmic transform of the cosmological density field shares the same key feature of the clipping transformation, in that it heavily penalises the highest density regions. However, the logarithmic transform also amplifies the influence of very low density regions, and as such is substantially more sensitive to shot noise than the clipping technique.  Furthermore, we find the power spectrum associated with the clipped field to be more straightforward to describe analytically, and we have presented the exact solution for the case of a Gaussian Random Field in (\ref{eq:clipAnalytic}). \citet{2011WangLog} developed an estimate for the power spectrum associated with the logarithm of the density field, and found that their prescription breaks down at approximately $k \sim 0.2 \hmpc$.

Further improvements may be possible using a combination of local and non-local transformations. For example, restoring the amplitude of the baryon acoustic oscillations \cite{EisSeoRecon} requires the reversal of large-scale motions, and so is not a suitable task for local transformations such as clipping. \citet{BOSS2012} recently applied a reconstruction technique to the CMASS sample from the Baryon Oscillation Spectroscopic Survey (BOSS). 

Another potential advantage of the clipping technique is that it conceals the regime in which baryonic physics most radically impacts the abundance of galaxies. The only information discarded by the clipping process - the total number of galaxies in high density environments - is not useful information as it is likely to be influenced by baryonic physics, which numerical simulations are unable to model.  Essentially, for the purpose of mitigating nonlinear effects, it appears beneficial to perform a cut not only at a particular wavenumber $\kmax$ but also at a maximum density $\rhomax$.

Future work will extend this study into redshift space, further develop models for the amplitude parameters $A_{ij}$, and apply this technique to data from galaxy redshift surveys. 

\noindent{\bf Acknowledgements}\\
The authors would like to thank Chris Blake, Michael Wilson, Benjamin Joachimi, Eric Linder, John Peacock and Martin White for helpful comments. FS is grateful for the hospitality of Swinburne University of Technology where part of this work was undertaken. FS and CH acknowledge support from the European Research Council under the EC FP7 grant number 240185. 

\appendix

\sec{The Power Spectra of Clipped Fields} \label{sec:app}


First we briefly review the derivation of the power spectrum for a symmetrically clipped Gaussian Random Field, as given by \citet{gross1994snr}, before generalising to the case of asymmetric clipping, higher order fields, and finally we briefly consider the behaviour of nonlinear galaxy bias when subject to clipping.

\subsection{Symmetric Clipping}

Clipping can arise in various communication systems, when a signal reaches the saturation level  \cite{van1966spectrum, hinich1967estimation, gross1994snr, li1998effects}.  \citet{gross1994snr} derive an exact expression for the spectral power of a Gaussian process with standard deviation $\sigma$ subject to a positive and negative threshold of equal magnitude, $\pm \delta_0$. To begin, Price's theorem relates the correlation function of the transformed field to the derivative of the transformation $g(\delta)$

\[
\frac{\partial \xi_c(r)}{\partial \xi(r)} = \left<   \frac{\partial g(\delta_1)}{\partial \delta_1} \frac{\partial g(\delta_2)}{\partial \delta_2}  \right>
\]
For the case of clipping the derivative of $g(\delta)$ is either unity or zero. The relevant expression is therefore

\[ \label{eq:clipDeriv}
\frac{\partial \xi_c(r)}{\partial \xi(r)} = \int_{-\delta_0}^{\delta_0}  \int_{-\delta_0}^{\delta_0} f(\delta_{1}, \delta_{2}, \rh) \ud \delta_1 \ud \delta_2
\]
where $\rh \equiv \xi(r)/\sigma^2 $ is the normalised correlation function, and the joint probability $f(\delta_{1}, \delta_{2}, \rh)$ of a GRF is given by
\[
f(\delta_{1}, \delta_{2}, r) = \frac{1}{2 \pi \sigma^2 \sqrt{1 - \rh^2}} \exp \left[  \frac{2 \rh \delta_{1} \delta_{2} - \delta_{1}^2  - \delta_{2}^2}{2 \sigma^2 \left[ 1 - \rh^2 \right]}  \right] 
\]
Applying Mehler's formula to the above expression leaves 

\[ \label{eq:Mehler}
\eqalign{
\frac{\partial \xi_c(r)}{\partial \xi(r)} = &\frac{1}{2 \pi \sigma^2} \sum_{n=0}^{\infty} \frac{\rh ^n}{2^n n!} \cr
& \left[   \int_{-\delta_0}^{\delta_0} \exp \left( \frac{-\delta_1^2}{2 \sigma^2} \right) H_n \left( \frac{\delta_1}{ \sqrt{2} \sigma} \right) \ud \delta_1 \right]^2
}
\]
which may be integrated to yield

\[  \label{eq:intSolution}
\frac{\partial \xi_c(r)}{\partial \xi(r)} =  \mathrm{erf}^2\left( u_0 \right)  +  \frac{4}{\pi} \sum_{n=2,4}^{\infty} \frac{\rh ^n}{2^n n!} e^{-2 u_0^2 } H^2_{n-1} \left( u_0 \right) 
\]
where the terms involving odd $n$ vanish, and we have defined the normalised threshold value $u_0 \equiv \delta_0/\sqrt{2} \sigma$. This leaves an expression for the correlation function of the clipped field

\[ 
\xi_c(r) =  \mathrm{erf}^2 \left( u_0 \right) \xi(r) + 4 \sigma^2 \sum_{n=1}^{\infty} C_{2n}(u_0) \left[ \rh \right]^{2n+1} \, .
\]
This may be transformed to the desired form 

\[ \label{eq:gross}
 P_c(k) =   \mathrm{erf}^2 \left(  u_0 \right) P(k)  +  4 \sigma^2 \sum_{n=1}^{\infty} C_{2n}(u_0)  \hat{P}^{*(2n+1)}(k)   \, ,
\]
where the distortion coefficient $C_n(x)$ is defined in (\ref{eq:cn}), $\hat{P}(k)$ is the normalised power spectrum, and the notation $\hat{P}^{*n}(k)$ represents a self-convolution of the order $n$.

\subsection{Asymmetric Clipping}

If only a positive threshold is applied then the lower integration limits in (\ref{eq:clipDeriv}) now extend to infinity

\[ \label{eq:clipOneSided}
\frac{\partial \xi_c(r)}{\partial \xi(r)} = \int_{-\infty}^{\delta_0}  \int_{-\infty}^{\delta_0} f(\delta_{1}, \delta_{2}, \rh) \ud \delta_1 \ud \delta_2  \, .
\]

One key difference from the symmetric case is that terms with odd $n$ no longer vanish due to the asymmetry of the integral. Proceeding as before, we arrive at 

\[
\eqalign{
 \xi_c(r) = \frac{1}{4} & \left[ 1 + \mathrm{erf}\left(u_0\right) \right]^2  \xi(r) \, + \cr
 & \frac{\sigma^2}{\pi} \sum_{n=1}^{\infty} \frac{\rh^{n+1}}{2^n (n+1)!} e^{-2 u_0^2} H^2_{n-1} (u_0) \, ,
 }
\]
which finally transforms to

\[ \label{eq:onesidedAppendix}
 P_c(k) = \frac{1}{4} \left[ 1 + \mathrm{erf}\left(u_0\right) \right]^2  P(k) + \sigma^2  \sum_{n=1}^{\infty} C_n (u_0) \hat{P}(k)^{*(n+1)}  \, .
\]

\subsection{Higher Order Fields}

Now consider a field which is the square of a GRF, defined by $e \equiv \delta^2 - \langle \delta^2 \rangle$ such that $\langle e \rangle = 0$. If we apply asymmetric clipping (involving only a positive threshold $\delta_0$) to a stand-alone square of a GRF, $\delta_G^2$, this is equivalent to performing symmetric (two-sided) clipping on $\delta_G$ at a threshold of $\delta'_0 = \pm \sqrt{\delta_0 + \langle \delta^2 \rangle}$.   The correlation function of this field is therefore given by

\[
\eqalign{
\xi_c(r, e) &\equiv \langle e_{c1}e_{c2} \rangle \cr
 &= \int_{-\infty}^{\infty}  \int_{-\infty}^{\infty} e_{c1}e_{c2} f(e_{c1}, e_{c2}) \ud e_{c1} \ud e_{c2} \cr
 &= \int_{-\delta'_0}^{\delta'_0}  \int_{-\delta'_0}^{\delta'_0}  \delta_1^2 \delta_2^2 f(\delta_{1}, \delta_{2}, \rh) \ud \delta_1 \ud \delta_2 \, + \cr
&  4 \delta_0'^2 \int_{-\delta'_0}^{\delta'_0}  \int_{\delta'_0}^{\infty}   \delta_2^2 f(\delta_{1}, \delta_{2}, \rh) \ud \delta_1 \ud \delta_2 \, .
}
\]
When the clipping is weak, $\delta_0>\sigma$, the second term is negligible. Decomposing into Hermite polynomials as before, and solving the leading order $n=0$ term leaves us with

\[
\xi_c(r, e) \simeq \frac{1}{\pi} \left[\sqrt{\pi} \mathrm{erf}(u'_0) - 2 u'_0 e^{-u_0'^{2}}     \right]^2  \xi(r,e) \, ,
\]
where 

\[
u'_0 \equiv \sqrt{\frac{\delta_0 + \sigma^2}{2 \sigma^2}} \, ,
\]
and $\sigma$ refers to the standard deviation of $\delta_G$, the underlying GRF. Equivalently
\[
P_c(k, e) \simeq \frac{1}{\pi} \left[\sqrt{\pi} \mathrm{erf}(u'_0) - 2 u'_0 e^{-u_0'^2}     \right]^2  P(k,e) \, .
\]
Using this leading order term alone typically recovers the clipped power spectrum to better than one per cent, provided the reduction in amplitude is not greater than a factor of two. Note that this function decays much more rapidly than the leading order term in (\ref{eq:onesidedAppendix}), demonstrating how clipping enhances the linear power spectrum relative to the higher order terms.

\subsection{Clipping Galaxy Bias} \label{sec:clipBias}
     
Consider a general form of local galaxy bias (e.g. \citet{fry1993biasing}; \citet{McDonald2006bias}; \citet{jeong2009perturbation})

\[
\delta_g (x) = \epsilon + c_1 \delta (x) + \frac{1}{2} c_2 \delta^2(x) + \frac{1}{6} c_3 \delta^3 (x) + \ldots 
\]
\vspace{1pt}

\noindent Truncating at third order in $\delta$ leads to the following expression for the power spectra, equation (10) from \cite{McDonald2006bias}

\[ \label{eq:mcdonald}
\eqalign{
P_g(k) &= P_0 + \left[c_1^2 + c_1 c_3 \sigma^2 + \frac{68}{21} c_1 c_2 \sigma^2 \right] P(k) \cr
 &+ c_1 c_2 P_{b2}(k)  + c_2^2 P_{b22}(k)  \, ,
}
\]
where $P(k)$ is the full (non-linear) matter power spectrum, while $P_{b2}(k)$  and $P_{b22}(k)$ are the lowest order deviations from linear bias.  After clipping, we should expect  the galaxy bias $b(k)$ to be linear at higher wave numbers than before. Terms in (\ref{eq:mcdonald}) involving $\sigma^2$, $P_{b2}(k)$, $P_{b22}(k)$ are suppressed relative to $c_1^2 P_L(k)$ in a similar manner to the one-loop terms, while the omitted higher order terms are subject to even stronger reductions in amplitude. We should therefore expect that even in the presence of significant nonlinear galaxy bias, the clipped galaxy power spectrum may be well described by

\[\label{eq:clipGalPk}
\eqalign{
P_{cg}(k) &= P_0 + A_{11} c_1^2 P_L(k) + A_{22} c_1^2 \left[ P_{22}(k) + P_{13}(k) \right] \cr
 &+ A_{b2} \left[c_1 c_2 P_{b2}(k)  + c_2^2 P_{b22}(k) \right]  \, .
}
\]
 
\bibliography{/Users/fergus/Gdrive/HomeSpace/Routines/dis}

\end{document}